\documentclass[openacc]{rstransa}

\def\IR{\mathbb{R}}\def\IZ{\mathbb{Z}}

\begin{document}

\title{Ramanujan's influence on string theory, black holes and moonshine}

\author{
Jeffrey A. Harvey$^{1}$}

\address{$^{1}$Enrico Fermi Institute and Department of Physics, University of Chicago, 5640 Ellis Ave., Chicago Illinois 60637, USA }

\subject{}

\keywords{}

\corres{Jeffrey A. Harvey\\
\email{j-harvey@uchicago.edu}}

\begin{abstract}
Ramanujan influenced many areas of mathematics, but his work on q-series, on the growth of coefficients of modular forms, and on mock modular forms stands out for its
depth and breadth of applications.  I will give a brief overview of how this part of Ramanujan's work has influenced physics with an emphasis on applications to string theory, counting of black hole
states and moonshine. This paper contains the material from my presentation at the meeting celebrating the centenary of Ramanujan's election as FRS and adds some additional material
on black hole entropy and the AdS/CFT correspondence.
\end{abstract}


\begin{fmtext}

\section{Historical remarks}

{\it My dream is that I will live to see the day when our young physicists, struggling to bring the predictions of superstring theory into correspondence with the facts of nature, will be led to enlarge their analytic machinery to include not only theta-functions but mock theta-functions \ldots But before this can happen, the purely mathematical exploration of the mock-modular forms and their mock-symmetries must be carried a great deal further. --Freeman Dyson} 

\end{fmtext}
\maketitle

It is a great pleasure and honor to be part of the celebration of the centenary of Ramanujan's election to Fellow of the Royal Society.  At the celebration of the centenary of Ramanujan's birth Freeman Dyson made the remarks quoted above. 
Since that time the mathematics of mock modular forms has in fact been carried a great deal further and some string theorists now include mock modular forms in their mathematical toolbox. Unfortunately it would be too much of a stretch to say that this has brought superstring theory into correspondence with the facts of nature. It has however improved our understanding of one of the best theoretical tests of string theory as a theory of quantum gravity, namely whether or not it provides a framework in which the entropy of black holes can be understood and computed.  In addition, other aspects of Ramanujan's work have also had an influence on developments in physics including in particular the structure of conformal field theory. They have also had great influence on the subject of moonshine, a still mysterious subject that seems to lie at the boundary of physics and mathematics.  I want to give a brief overview of how Ramanujan's ideas have influenced these subjects.

The general topic of this talk is Ramanujan's influence on physics and physicists and in particular his influence on topics connected to string theory. Since I am at the University of Chicago
it seems appropriate to first mention Ramanujan's influence on the Noble laureate S. Chandrasekhar who spent most of his career at the University of Chicago. At Hardy's request Chandrasekhar tracked down the passport photograph of Ramanujan that we have all seen. This happened on a trip to India in 1936 which is described in rich detail in \cite{wali}.  Chandrasekhar and Ramanujan had  a common social background and it is clear that Chandrasekhar had great admiration for Ramanujan and his work. As testimony to this Chandrasekhar played an important role in
commissioning a bust of Ramanujan. He and his wife acquired two of the original four busts and one of these was then donated to the Royal Society
as described in \cite{chandraI}.  As I hope I will make clear, Ramanujan's influence on physics has grown greatly since that time and there are many
hints that his influence will become even greater in the future.

\section{Partitions and the Hagedorn Temperature}

Let me know turn to some physical considerations. 
Quantum systems at finite temperature are described by a thermal partition function
\begin{equation}
Z(\beta) = {\rm Tr}_{\cal H} e^{-\beta H} = \sum_E d(E) e^{-\beta E}
\end{equation}
where ${\cal H}$ is the Hilbert space of the system, $\beta=1/kT$ is the inverse of the temperature times Boltzmann's constant and $H$ is the Hamiltonian operator of the
system whose eigenvalues are the allowed energies $E$. The sum over energies is in general both a sum over the discrete spectrum and an integral over the continuous part of the energy spectrum and $d(E)$ is the
number of states of energy $E$ or density of states if we are considering the continuous part of the spectrum. If $d(E) \sim e^{a E}$  with $a>0$ at large energies $E$ then there is a limiting temperature since when $\beta<a$ the partition function diverges.
This kind of growth in the density of states with energy does not occur in typical quantum systems but it does appear in string theory and in the physics of black holes, and as I will
explain, is connected to the work of Hardy-Ramanujan on the growth of the number of partitions of integers $p(n)$.
\footnote{I thank John Keating for pointing out work on the energy levels of heavy nuclei which provided an early application of the Hardy-Ramanujan asymptotic formula for $p(n)$ to a physical system. See \cite{bohr,vanlier} for further details. }

The fundamental reason why Ramanujan's work has had a significant impact on physics is due to the fact that modular forms play a central role in string theory and two-dimensional conformal
field theory (CFT). The basic reason for this is the following.  A closed string at a fixed time is an $S^1$ embedded into some ambient space. Finite temperature in quantum systems can be described
in a path integral formalism in terms of a periodic Euclidean time $t_E= -it$ with $t_E \simeq t_E+\beta$. We are thus led to consider path integrals over a two-dimensional Euclidean space with
topology $S^1 \times S^1$ or more generally, since the theory is conformally invariant, we have path integrals that depend on the conformal equivalence class of a two-torus or elliptic curve $E_\tau$
labelled by $\tau$ in the upper half plane and invariant under the action of the modular group $PSL(2, \IZ)$ which acts as global diffeomorphisms of the two-torus.

Now let's consider the various modes of a string.  For simplicity we will consider an open string with fixed endpoints, the generalization to closed
strings with spatial world-volume $S^1$ is straightforward.  A string with fixed endpoints has a lowest mode with no nodes, a first excited mode with
one node and so on. For example, if the lowest mode of a piano string is middle $C$, the next mode with one node is an octave above middle $C$, the 
next mode is the $G$ a fifth above this $C$ and so on. In a quantum theory each mode is created by a creation operator $a^\dagger_{-n}$ which creates
the $n$th mode with energy $E =E_0 \sqrt{n}$ for some constant $E_0$ when acting on the vacuum state $0 \rangle$ with no excitations (more precisely $E$ is the mass of the corresponding spacetime state of the string). Thus there are two states with energy $E_0 \sqrt{2}$, the states $a^\dagger_{-2} |0 \rangle $ and $ a^\dagger_{-1} a^\dagger_{-1} |0 \rangle$ and there are $p(n)$
states with energy $E_0 \sqrt{n}$ where $p(n)$ is the partition number of $n$, that is the number of ways of writing $n$ as a sum of natural numbers. 
To be a bit more concrete about the relation between $p(n)$ and string theory, the generating function for $p(n)$ is
\begin{equation}
\frac{1}{\prod_{m=1}^\infty(1-q^m)} = \sum_{n=0}^\infty p(n) q^n
\end{equation}
while the partition function of the open bosonic string is the inverse of the famous weight $12$ cusp form
\begin{equation}
Z_{open}^{bos}(q) = \frac{1}{\eta(q)^{24}} =\frac{(q^{-1/24})^{24} }{\prod_{m=1}^\infty (1-q^m)^{24}} \, .
\end{equation}
The additional factor of $q^{-1/24}$ associated to each of the $24$ independent oscillators (corresponding to the 24 directions transverse to a string
in $26$ spacetime dimensions) can be interpreted in string theory as a result of regularizing the infinite zero point energy $(1/2) \sum_n n$ of an infinite number of harmonic oscillators. In a mathematical context this result was presented by Ramanujan in his second letter to Hardy  where he wrote ``I told him that the sum of an infinite no. of terms of the series: $1+2+3+4 + \cdots = -1/12$ under my theory \ldots I dilate on this simply to convince you that you will not be able to follow my methods of proof  if I indicate the lines on which I proceed in a single letter."  This is certainly one of the most cheeky explanations ever offered for a lack of mathematical proof. 

The famous result of Hardy-Ramanujan \cite{hardyram} that
\begin{equation}
p(n) \sim \frac{1}{4 n \sqrt{3}} \exp \left( \pi \sqrt{ \frac{2n}{3}} \right)
\end{equation}
then shows that we are in the situation described earlier where the density of states grows exponentially with energy and this implies that
string theory has a limiting temperature known as the Hagedorn temperature. This behavior also has consequences in the real world since there
is good evidence that there is an effective string theory description of the strong interactions that binds quarks into hadrons. In that context the
limiting Hagedorn temperature is around $10^7$ degrees Kelvin and indicates a transition to a description in terms of free quarks and gluons rather
than their string-like hadronic bound states.

\section{Modular functions and BPS state counting}

In supersymmetric quantum field theories and string theories a special role is played by what are called BPS states. The initials BPS stand for Bogomolny, Prasad and Sommerfeld who
did early work analyzing equations for soliton states that were later understood to be connected to supersymmetry. 
BPS states are in special, ``short" representations of the supersymmetry algebra due to the fact that some of the supercharges of the theory
annihilate these states,
\begin{equation}
Q^\alpha |BPS \rangle=0 \, .
\end{equation}

One of the simplest contexts where counting half BPS states (meaning they preserve half of the supercharges) plays an important role is  heterotic
string theory on a six-dimensional spacetime torus. In this case the BPS counting function is the inverse of the famous weight $12$ cusp form, 
\begin{equation}
\frac{1}{\Delta(\tau)} = \frac{1}{\eta(\tau)^{24}} \, .
\end{equation}

 In more complicated situations we have path integrals depending on $\tau$ and point on the elliptic curve $z$, or path integrals involving genus two Riemann surfaces. These lead to BPS state counting functions that involve Jacobi forms and Siegel forms respectively. For example, a BPS counting
function that counts $1/8$ BPS states in type II string theory on a six-dimensional spacetime torus is the Jacobi form
\begin{equation}
\varphi_{-2,1}(\tau,z)= \frac{\theta_1(\tau,z)^2}{\eta(\tau)^6} \, .
\end{equation}

BPS states in type II string theory on a product of a $K3$ torus and a two-dimensional spacetime torus have a BPS counting function given
by $1/\Phi_{10}(\tau,z,\sigma)$
where $\Phi_{10}$ is the weight ten Siegel cusp form
\begin{equation}
\Phi_{10}(\tau,z,\sigma)= qyp \prod_{(r,s,t)>0} (1-q^s y^t p^r)^{2c(4rs-t^2)} 
\end{equation}
where  $q=e^{2 \pi i \tau}$, $y=e^{2 \pi i z}$, $p=e^{2 \pi i \sigma}$. 
In this expression the coefficients $c(m)$ are the coefficients of a Jacobi form of weight zero and index one that also appears as a BPS counting
function and is proportional to the elliptic genus of a $K3$ surface,
\begin{equation}
Z_{ell}(\tau,z;K3)=2 \varphi_{0,1}(\tau,z)= 2 \sum_{n,r} c(4n-r^2) q^n y^r \, .
\end{equation}

This relation between the coefficients of a Jacobi form and those of a Siegel form is an example of a multiplicative or Borcherds lifts and such
relations have been much studied in pure mathematics. But such relations often have a physical interpretation as well. In this case the Siegel form
$\Phi_{10}$ is closely related to the generating function of the elliptic genus of the symmetric product orbifold \cite{Dijkgraaf}
\begin{equation}
S^N K3 = K3^{\otimes N}/S_N \, .
\end{equation}
It is natural to wonder why $K3$ surfaces show up in these counting problems. The answer is that $K3$ is the simplest non-trivial Calabi-Yau
space and thus it provides both a consistent background for studying supersymmetric string theory and a background that is simple enough that the
problem of computing the entropy of black holes can be solved in great detail. 

\section{The Mystery of Moonshine}

There are special classes of modular and mock modular forms that have integer coefficients in their $q$ expansion. These forms were of particular interest to Ramanujan and of course play a central role in number theory.  In a number of interesting cases  such (mock) modular forms have a secret life: their coefficients are characters of finite groups. In some sporadic, exceptional and special cases, often linked to genus zero subgroups of $SL(2, \mathbb{R})$, this connection is referred to as moonshine.  While there is no unified understanding of all examples that have been found of moonshine, there is a principle that has proved useful in looking for new types of moonshine and that is the principle of minimal exponential growth of coefficients.
All of the BPS counting functions we have encountered have negative powers of $q$ in the Fourier expansion. They thus diverge as $\tau \rightarrow \infty$, that is at the cusp at infinity, and Hardy-Ramanujan type estimates of the growth of coefficients tell us that they have exponential growth.
The principle says we should look for moonshine amongst forms of a given type with the smallest possible exponential growth. 

The classical example of moonshine involves the weight zero modular $J$ function which is invariant under the action of $SL(2, \mathbb{Z})$. 
A basis for weight zero modular functions that diverge only at the cusp at infinity can be constructed by starting with the constant function
and $J(\tau)$ and then taking powers of $J$ and subtracting off smaller powers so that the leading terms in the $q$ expansion has the form
$q^{-m}+O(q)$. This gives the following set of functions (or rather their $q$ expansions:
\begin{align}
\begin{split}
J_0 &= 1 \\
J_1 &= q^{-1}+196884 q +21493760 q^2 +  \cdots \\
J_2 &= q^{-2}+42987520 q + \cdots \\
 .& \\
J_m &= q^{-m} + O(q) = \sum_{n=-m} c_m(n) q^n
\end{split}
\end{align}

The circle method gives
\begin{equation}
c_m(n) \sim \exp(4 \pi \sqrt{mn})
\end{equation}
and so $J_1(\tau)=J(\tau)$ is the weight zero modular function of minimal exponential growth. 

Monstrous Moonshine started with McKay's observation that
\begin{equation}
196884=196883+1
\end{equation}
expresses the first nontrivial coefficient of $J(\tau)$ in terms of the sum of dimensions of the two smallest irreducible representations of the Monster
group. J. Thompson generalized this observation to 
\begin{equation}
21493760 = 21296876+196883+1
\end{equation}
and suggested that the coefficients $c_1(n)$ should be interpreted as the dimensions of vector spaces $V_n$ providing modules for the Monster group. 
\begin{equation}
c_1(n)= {\rm dim} V_n = {\rm Tr}(1)|_{V_n}\, .
\end{equation}
This naturally leads one to consider the McKay-Thompson series
\begin{equation}
T_g(\tau)= \sum_n {\rm Tr}(g)|_{V_n} q^n \, .
\end{equation}
Note that $T_g(\tau)$ depends only on the conjugacy class of $g$, so there are in principle 194 different functions $T_g(\tau)$. Conway and Norton
computed these functions and made a remarkable conjecture concerning their structure. The modular function $J(\tau)$ is a hauptmodul, or generator
of the function field of weight zero functions invariant under $SL(2, \IZ)$. The upper half plane ${\cal H}$ is acted on by $PSL(2,\IZ)$ and the quotient
$PSL(2,\IZ) \backslash {\cal H}$ with the addition of the cusp at infinity is mapped by $J(\tau)$ to the Riemann sphere. Thus
$PSL(2,\IZ)\backslash {\cal H} \cup {i \infty}$  is a genus zero Riemann surface. Conway and Norton conjectured that the functions $T_g(\tau)$ are also hauptmoduls, playing a role analogous to that
of $J(\tau)$ but for genus zero group $\Gamma_g$ arising from a group between $\Gamma_0(N)$ and its normalizer in $PSL(2, \IR)$.These
observations were partly explained by the construction by Frenkel, Lepowsky and Meurman  \cite{flm} of a holomorphic conformal field theory or
vertex operator algebra in mathematical language with the monster sporadic group as automorphism group and with a one-loop partition function
given by $J(\tau)$. This culminated in the proof of the genus zero property by Borcherds \cite{borch}.

Other more recent examples of moonshine are also linked to modular forms of minimal exponential growth. For example, in a study of Borcherds
lifts Borcherds and Zagier introduced and studied a basis for  weakly holomorphic weight $1/2$ forms invariant under $\Gamma_0(4)$ and
in the Kohnen plus space \cite{borch, zag}. The $q$ expansions of the first few basis forms are
\begin{align}
\begin{split}
f_0 &=\theta(\tau) = 1+2 q + 2 q^4 + 2 q^9 + 2 q^{16} + \cdots \\
f_3 &= q^{-3} -248q +26752 q^4- 85995 q^5+1707264 q^8 + \cdots \\
f_4 &= q^{-4}+ 492 q +143376 q^4 + 565760 q^9 + \cdots
\end{split}
\end{align}
Here the Jacobi theta function plays the role of the constant function with coefficients $c(n)$ that are constant with $n$ and the function with
slowest exponential growth is $f_3(\tau)$ plus any multiple of $\theta(\tau)$. Following \cite{hr} we note that the combination with vanishing term
linear in $q$,
\begin{equation}
2 f_3 + 248 \theta(\tau) = 2 q^{-3} + 248 + 2 \times 27000 q^4 - 2 \times 85995 q^5 + \cdots
\end{equation}
has coefficients which involve $248, 27000, 85995$, all of which are dimensions of irreducible representations of the Thompson sporadic group. 
Specific conjectures regarding this new kind of Thompson moonshine were made in \cite{hr} and these conjectures were later proved in \cite{gm}. 

An example of moonshine with a closer link to the work of Ramanujan emerged from the work of Eguchi, Ooguri and Tachikawa \cite{eot}. 
String propagation on a $K3$ surface defines a $N=4$ superconformal invariant two-dimensional quantum field theory. On general grounds
the elliptic genus of a $K3$ surface must then have a decomposition into characters of the $N=4$ superconformal algebra. This algebra has two classes of representations sometimes referred to as BPS and non-BPS. In \cite{eot} they found that the multiplicities of the non-BPS representations
are the coefficients of a weight $1/2$ mock modular form with $q$ expansion
\begin{equation}
H^{(2)}(\tau) = q^{-1/8} \sum_n a(n) q^n =2 q^{-1/8} \left( -1 + 45 q + 231 q^2 + 770 q^3 + \cdots \right)
\end{equation}
and they pointed out that the coefficients $45$, $231$, $770$ are dimensions of irreducible representations of the sporadic Mathieu group
$M_{24}$.  This led to a flurry of work to determine the (mock) modular structure of the associated McKay-Thompson series \cite{cheng2010, Gaberdiel:2010ch, Gaberdiel:2010ca, Eguchi:2010fg}.  What I mean by this is that the observations relating the coefficients $a(n)$ of the mock modular form $H^{(2)}$ to the representation theory of $M_{24}$ suggests that there is an  infinite-dimensional, graded $M_{24}$ module 
\begin{equation}
K^{(2)} = \oplus_{k \in \mathbb{Z}} K^{(2)}_{k-1/8}
\end{equation}
such that $a(k) = {\rm dim} (K^{(2)}_{k-1/8}$ for $k>0$.  For each conjugacy class $[g]$ of $M_{24}$ we can then define a McKay-Thompson series
\begin{equation}
K^{(2)}_{[g]} (\tau) = -2 q^{-1/8} + \sum_{k =1}^\infty {\rm tr}_{K^{(2)}_{k-1/8}}(g) q^{k-1/8} \, .
\end{equation}
The work cited above conjectured the existence of $K^{(2)}$ such that the $K^{(2)}_{[g]}$ have specified mock modular properties, essentially that
they are mock modular forms of level $o[g]$. The existence of $K^{(2)}$ with these properties was then proved by Gannon \cite{gannon}.

In Umbral Moonshine \cite{um, umnl} Mathieu Moonshine is expanded to a much larger class of mock modular forms and groups and a new element enters the picture: the $23$ even, self-dual, rank $24$ lattices with non-zero root systems known as Niemeier lattices. We let $X$ denote the root system of the Niemeier lattice and label the lattice by its (unique) root system as $L^X$.  Niemeier lattice $L^X$ determines a finite group as the quotient of the automorphism group of $L^X$ by its normal subgroup ${\rm Weyl}(X)$ generated by reflections in the roots of $X$:
\begin{equation}
G^X = {\rm Aut}(L^X)/{\rm Weyl}(X)
\end{equation}
It turns out that one can associate a vector-valued mock modular form $H^{X}$ to each $X$ that exhibits moonshine for $G^X$ meaning that again
there is a conjectured infinite dimension module $K^X$ and McKay-Thompson series $H^X_{[g]}$ with specified mock modular properties. The existence of such modules was proved in \cite{dgo}. Furthermore, the eigenvalues of the Coxeter element of each $X$ can be used
to construct a ratio of $\eta$ functions that is a hauptmodul for a genus zero subgroup of $SL(2, \mathbb{R})$. This connection was significantly strengthened in work by Cheng and Duncan \cite{Cheng:2016klu}
that generalized this to give a genus zero classification of what they called optimal mock Jacobi theta functions and showed that these include as their theta coefficients all the mock modular forms
of umbral moonshine. 

To give a taste of how some of the most famous of Ramanujan's mock modular functions appear in umbral moonshine consider his third
order mock theta function
\begin{equation}
f(q) =\sum_{n=0}^\infty \frac{q^{n^2}}{(1+q)^2 \cdots (1+q^n)^2} = 1+q-2 q^2 + 3 q^3 -3 q^4 +3 q^5 - 5 q^6 + \cdots
\end{equation}

In the $X=A_2^{12}$ case of umbral moonshine a weight $1/2$, two-component mock modular form $H^{A_2^{12}}$ appears and its first component has a $q$ expansion given by 
\begin{equation}
H^{A_2^{12}}_1 = 2 q^{-1/12} \left( -1 + 16 q + 55 q^2 + 144 q^3 + 330 q^4 + 704 q^5 + 1397 q^6 + \cdots \right)
\end{equation}
Conjecture 6.1 of \cite{umnl}  implies that there is a natural $2.M_{12}$ super-module 
\begin{equation}
K_1^{A_2^{12}} = \bigoplus_{\substack{D \in \mathbb{Z}  \\ D=1 ~{\rm mod}~12}} K^{(3,1)}_{1,-D/12}
\end{equation}
such that the graded super-character attached to an element $g \in 2.M_{12}$ is the first component of a vector valued mock modular form
\begin{equation}
H^{A_2^{12}}_g(\tau) = \sum_{\substack{D \in \mathbb{Z}  \\ D=1 ~{\rm mod}~12}} {\rm str}_{K^{A_2^{12}}_{1,-D/12}}(g) q^{-D/12}
\end{equation}
with specified shadow and modular properties that can be found in \cite{umnl}. 
Given that $1$, $16$, $55$ and $144$ are all dimensions of irreducible representations of $2.M_{12}$ it is perhaps not surprising that the associated representations appear as the first terms
: $K^{A_2^{12}}_{1,-1/12}= -2 V_1$, $K^{A_2^{12}}_{1,11/12}= V_4+V_5$, $K^{A_2^{12}}_{1,23/12}= 2 V_9$ and so on. The first few terms in $H^{A_2^{12}}_{[g]}$ for $[g]=2B$ can then be computed from the
character table of $2.M_{12}$, part of which is given in Table 1 and one finds that they agree with the $q$ expansion of $-2 q^{-1/12} f(q^2)$. This agreement was conjectured and then proved in \cite{dgo}
to hold to all orders in the $q$ expansion. So we see that Ramanujan's mock modular forms have a great deal of structure in them that was not at all obvious when they were first constructed.
They are linked both to genus zero subgroups of $SL(2, \mathbb{R})$ and to characters of finite groups closely related to sporadic groups, in this example the sporadic group $M_{12}$.

\begin{table}[!h]
\caption{Part of the character table of $2.M_{12}$}
\label{char_tab}
\begin{tabular}{|c|cccc|}
\hline
$[g]$ & 1A & 2A & 4A & 2B \\
\hline
$\chi_1$ & 1 & 1 & 1 & 1\\
$\chi_2$ & 11 & 11 & -1 & 3 \\
$\chi_3 $ &    11     & 11    &  -1    &  3  \\
$\chi_4 $ &  16       & 16    &   4   &  0  \\
$\chi_5 $ &    16     & 16    & 4     &  0  \\
$\chi_6$ &   45      & 45    &  5    & -3   \\
$\chi_7$&  54       &  54   &   6   & 6   \\
$\chi_8 $ &   55      &  55   &  -5    &  7  \\
$\chi_9$ &  55       &   55  &  -5    & 1   \\
$\chi_{10}$ &    55     &   55  &  -5    &  1  \\
\hline
\end{tabular}
\vspace*{-4pt}
\end{table}

\section{Mock  Modular Forms and Black Hole Entropy}

One of the compelling reasons to think that string theory provides us with a consistent theory of quantum gravity is that it provides us with very
precise ways to compute the entropy of a large class of black holes. Following on earlier studies of properties of black holes,  work of Bekenstein and Hawking led to the idea that black holes are thermodynamic objects, characterized by an entropy $S$, and a temperature $T$ that characterizes the thermal blackbody radiation emitted by the black hole due to quantum effects. The temperature and entropy are determined in terms of the mass $M$ of the black hole and the area $A$ of its event horizon by
\begin{equation}
S=\frac{A}{4}, \qquad T= \frac{1}{8 \pi M}
\end{equation}
where as usual I have used natural units with $\hbar =c=G=1$. 
Now these are essentially thermodynamic results, computed from the classical geometry of the black hole,  but we know in ordinary physical systems that thermodynamics can be derived from statistical mechanics and in statistical mechanics the entropy
is computed as the logarithm of the number of quantum states of the system that are consistent with the given macroscopic parameters. In the simplest case of a neutral black hole that is not rotating
we should then have $S= \ln d(M)$ where $d(M)$ is the number of quantum states of the black hole consistent with it having mass $M$. This can be generalized to include black holes with charge
$Q$ and angular momentum $J$ in which case one would like to compute $d(M,J,Q)$ and check whether its logarithm agrees with the entropy computed using the black hole geometry.  Starting with work
of Strominger and Vafa a method of computing black hole entropy in string theory has been developed and applied to increasingly complicated situations. 
The basic idea is to construct a two-dimensional conformal field theory that describes string propagation on a configuration of D-branes whose mass, 
charge and angular momentum match that of the desired black hole and such that the configuration (and black hole) is BPS, or near BPS. The
number of configurations of given mass, charge and angular momentum can then be determined in terms of the asymptotic growth of the
partition function or related quantities in the conformal field theory. In the physics literature this is referred to as using the Cardy formula, but it
is closely connected to the methods of Hardy-Ramanujan.  The final step in the argument involves taking a strong coupling limit in which the configuration of D-branes goes over to a black hole that has the same conserved quantum numbers as the D-brane configuration. 

In more recent work on counting black hole states mock modular forms have also made an appearance. This started with the papers
\cite{fareytail,manmoore} and is presented in a particularly clear form aimed at both physicists and mathematicians in Dabholkar, Murthy and Zagier \cite{dmz}. The basic problem these papers address is that in the strong coupling limit the D-brane configuration does not necessarily go over
to a single-centered black hole. Rather, there can also be multi-centered configurations of black holes \footnote{Because the black holes have both mass
and electric and magnetic charge they are not necessarily attracted to each other, a phenomenon which is common for BPS states.} and the degeneracies of these multi-centered configurations can jump in crossing walls of marginal stability in the moduli space of the theory. Since one wants to compute
the degeneracy of single center black holes in order to compare to the Bekenstein-Hawking formula it is necessary to subtract off these multi-center
black holes. Remarkably, once one does that one is left with a counting function for single center black holes that is a mock modular form rather than a
modular form.

\section{Black Hole Entropy, the AdS/CFT Correspondence and Growth of Coefficients of Families of Modular Functions}

I would like to end with a brief discussion of a topic that I did not have time for in my presentation but that has been a topic of some interest in the string theory community and may also be of interest to some mathematicians. It is connected to what is known as the $AdS_3/CFT_2$ correspondence. There is now extremely good evidence, although no rigorous mathematical proof, that certain two-dimensional conformal field theories have a dual description in terms of string theories or gravity theories in three-dimensional Anti-DeSitter space. Of particular interest are CFTs whose dual description can be approximated by gravitational theories in $AdS_3$ rather than string theories, that is they can be described approximately by the low-energy limit of string theory. For this to be the case that radius of $AdS_3$ must be much larger than the Planck scale (or more precisely the string scale) where string
theory effects become important. Since AdS/CFT duality relates the radius $ \ell$ of $AdS_3$ to the central charge of the CFT via the relation
\begin{equation}
c =\frac{3 \ell}{2G}
\end{equation}
where $G$ is the three-dimensional Newton's constant, large $\ell$ compared to the Planck length requires large central charge $c$.

Recently some other constraints on two-dimensional CFTs with gravity duals have been discussed that result from demanding that the CFT
partition function mirror a fairly well understood property of quantum gravity in AdS space, namely that there should be a phase transition as a function of temperature, known
as the Hawking-Page transition, where the partition function changes from being dominated by a thermal gas of gravitons in AdS space at low temperatures to a black hole in AdS space at high temperature.  The partition function of three-dimensional gravity can, in a semi-classical expansion, be written as a sum over saddle points.  The solution corresponding to a thermal gas has action $\log Z_{\rm therm}= c \beta/12$ at inverse temperature $\beta$ while the black hole solution has action $\log S_{BH} = \pi^2 c/3 \beta$. This leads to a phase transition at $\beta=2 \pi$. A dual CFT should
exhibit a similar phase transition as a function of temperature.

In a two-dimensional conformal field theory one has two copies of the Virasoro algebra with modes $L_n$ and $\bar L_n$, $n \in \mathbb{Z}$ and with in principle different
central charges $c, \bar c$. For simplicity I will assume that $c = \bar c$. States in the Hilbert space ${\cal H}$ of the CFT can be labelled by their conformal weights $h, \bar h$
which are their eigenvalues under the action of $L_0$ and $\bar L_0$. The partition function of the theory is
\begin{equation}
Z(\tau)= {\rm Tr} q^{L_0-c/24} {\bar q}^{ \bar L_0 - c/24}
\end{equation}
where $q=e^{2 \pi i \tau}$, $\bar q= e^{-2 \pi i \bar \tau}$ and $\tau$ is a complex number lying in the upper half plane ${\rm Im} \tau >0$. It is important that in general
$Z$ is neither holomorphic nor anti-holomorphic in $\tau$. Physicists often denote this by writing $Z(\tau, \bar \tau)$. 

To connect the partition function above to our earlier discussion we write $\tau= (\theta+ i \beta)/2 \pi$ and for now set $\theta=0$. We also denote the
$L_0$ and $\bar L_0$ eigenvalues by $h, \bar h$ and define $E=h+ \bar h - c/12$. Then the partition function becomes
\begin{equation}
Z(\beta)= \sum_E e^{-\beta E}
\end{equation}
so that the CFT partition function just becomes the thermal partition function described earlier, albeit for a special kind of two-dimensional
quantum field theory. 
Recall that the thermodynamic energy $E(\beta)$ and entropy $S(\beta)$ can be computed from the partition function as
\begin{align}
\begin{split}
E(\beta) &= - \partial_\beta Z(\beta),  \\
S(\beta) &= (1- \beta \partial_\beta) Z(\beta) \, .
\end{split}
\end{align}

The partition function of a CFT is given by a path integral over a two-dimensional torus or elliptic curve with modular parameter $\tau$ and
should be invariant under modular transformations. The $S$ modular transformation taking $\tau \rightarrow -1/\tau$ takes $\beta \rightarrow 4 \pi^2/\beta$. We therefore should demand that $Z(\beta)= Z(4 \pi^2/\beta)$. Modular invariance as in the Hardy-Ramanujan analysis can be used
to estimate the growth of states with energy. One finds that at fixed $c$ and $E \rightarrow \infty$ that the growth of states, that is the entropy, behaves
in a way first analyzed by Cardy\cite{cardy}
\begin{equation}
S_{Cardy} = 4 \pi \sqrt{\frac{c E}{6}}
\end{equation}
This formula is commonly used to determine the entropy of black holes in the dual gravity theory. However, strictly speaking one should analyze
the entropy not as $ E \rightarrow \infty$ with $c$ fixed, but rather in the limit that $ c \rightarrow \infty$ with $E \sim c$.  The paper \cite{Hartman:2014oaa} analyed the conditions on a CFT in order that the same expression holds for the entropy in this limit of large $c$ and $E$ and that
there is a universal phase transition analogous to the Hawking-Page transition in gravity. Their result was that if the spectrum of the theory
is divided up into a light spectrum of states with $-c/12 \le E \le \epsilon$ with $\epsilon$ a positive quantity that goes to zero as $c \rightarrow \infty$
and a heavier spectrum with $ E \ge \epsilon$ 
then the entropy formula of Cardy holds provided that the light density of states obeys
\begin{equation}
\rho(E) \le e^{2 \pi (E+c/12}, ~~ E \le \epsilon
\end{equation}
and that one then finds a phase transition at large $c$ with
\begin{equation}
\log Z(\beta) = \frac{c}{12} {\rm max}~(\beta, 4 \pi^2/\beta)
\end{equation}
indicating a phase transition at $\beta= 2 \pi$.  Other consequences of the matching to 3d gravity have been explored more recently, for example
in \cite{Benjamin} constraints on the elliptic genera of CFT with $N=2$ supersymmetry were derived using related conditions. 

In general the partition functions of these CFTs are not holomorphic or even weakly holomorphic functions of $\tau$ and there is
little connection to the holomorphic or weakly holomorphic modular forms and functions that are mainly studied by mathematicians.
However there are certain CFTs known as Rational Conformal Field Theories (RCFTs) that have a much closer connection to modular
functions. In these theories there is a chiral algebra which contains the Virasoro algebra such that the partition function factorizes into a finite sum of the form
\begin{equation}
Z= \sum_{i, \bar j} N_{i, \bar j} \chi_i (\tau)  \bar \chi_{\bar j}(\tau)
\end{equation}
where the indices $i, \bar j$ run over a finite set of indices labelling irreducible representations $V_i$ of the chiral algebra, $N_{i, \bar j} \in \mathbb{Z}$ and 
\begin{equation}
\chi_i = {\rm Tr}_{V_i} q^{L_0 - c/24} \, .
\end{equation}
There is also an anti-chiral algebra which for simplicity we assume is isomorphic to the chiral algebra. A famous RCFT connected to moonshine
is the Monster CFT with partition function
\begin{equation}
Z = |J(\tau)|^2 \, .
\end{equation}
In \cite{witten3d} it was conjectured that this was the first of a set of extremal CFTs with central charges $c=24k$ and completely factorized
partition functions. Unfortunately the evidence of the existence of such a sequence of CFTs is very mixed.  RCFTs with only two independent
characters include an example whose characters are well known to number theorists. The so-called Yang-Lee model is a $c=-22/5$ RCFT
with characters
\begin{align}
\begin{split}
\chi_0 & = q^{-1/60} G(q) \\
\chi_{1/5} &= q^{11/60} H(q)
\end{split}
\end{align}
where $G(q)$ and $H(q)$ are the famous $q$-series of the original Rogers-Ramanujan identities.  One can construct a sequence of possible
characters of two character RCFTs with increasing $c$ using an adaptation of Hecke operators to the characters of RCFT \cite{harwu} but it
is not known whether there are actual RCFTs with these characters.

There is a certain amount of tension between known sequences of CFTs that obey the conditions required to have a gravity dual and RCFTs that
remain rational in the large $c$ limit. I have actually not defined precisely what this means, since clearly the size of the chiral algebra would also
have to increase in the large $c$ limit, but from a practical point of view a partition function which is a finite sum of products of holomorphic and anti-holomorphic terms in the large $c$ limit would be a strong indication of the desired behavior. It would be very interesting to exhibit RCFTs with gravity duals since these are likely to correspond to especially interesting classes of three-dimensional gravity theories and thus might teach us something interesting about quantum gravity, including whether
there is the sort of connection hinted at here and elsewhere between moonshine and three-dimensional gravity.

\section{Conclusion and Speculation}

I have tried to give a brief overview of several topics connected to string theory and physics where Ramanujan's work on modular forms
and mock modular forms has had an influence, namely special conformal field theories connected to moonshine and understanding the
counting of BPS states and the entropy of black holes in string theory. At first glance these two topics have little in common other than that they both utilize some of the mathematics developed by Ramanujan.  There are however speculative ideas about how these may eventually be connected.
One set of ideas runs roughly as follows. The data of a two-dimensional conformal field theory (or holomorphic VOA) includes specification of its
central charge $c$, a quantity that appears as a central term in the Virasoro algebra. The CFTs that are currently connected to moonshine
phenomena have specific, small values of $c$, $c=24$ for the Monster CFT, $c=12$ for a VOA exhibiting moonshine for the Conway group $Co_1$, and $c=6$ for the $K3$ CFT related to Mathieu Moonshine.
On the other hand, the CFT that appear in black hole counting problems, or rather the families of CFTs, have central charges $c(N)$ with
$c(N) \rightarrow \infty$ as $N \rightarrow \infty$. CFTs with large $c$  and obeying the conditions discussed in the previous section are expected to have a dual description in terms of three-dimensional gravity in hyperbolic space $\mathbb{H}_3$ (in the case of Euclidean signature) or Euclidean $AdS_3$ as it is known in the physics literature.  The growth of coefficients of the partition functions of CFTs via this correspondence end up being related to the entropy of black hole solutions in $AdS_3$.  So at large $c$ we have a fairly well understood, at least at a physical level of rigor, relation between two-dimensional conformal field theories and black holes in three-dimensional gravity.  We also know there are small central charge CFTs with special properties, that is they are connected to sporadic groups via moonshine. It is natural to wonder whether there might also be a three-dimensional gravity version of moonshine phenomenon when the three-dimensional spaces are highly curved and thus not easily described just by Einstein gravity. The idea of a connection between moonshine and three-dimensional gravity has been discussed starting with \cite{witten3d}. One common feature of the modular functions and Jacobi forms appearing in both moonshine and 3d gravity is that their coefficients can be expressed in terms of Rademacher series. These series of course are closely connected to Ramanujan's work with Hardy on the growth of the number of partitions $p(n)$ \cite{hardyram} and the further developments due to Rademacher that gave a convergent expression for $p(n)$ \cite{radI} and then later a series expansion of the coefficients of the modular $J$ function through a regularized version of Poincar{\'e} sum \cite{radII}.  These sums have a geometric interpretation in three-dimensional gravity that was developed in \cite{fareytail,manmoore, fd} and has been subsequently used to derive results for black hole entropy via supersymmetric localization in supergravity theories. See for example \cite{Dabholkar:2011ec, Dabholkar:2014ema}. One might hope that this common mathematical ingredient is not just a coincidence but rather points towards an eventual connection between moonshine and quantum gravity that would provide new insight into both topics.

\vskip6pt

\enlargethispage{20pt}

\funding{This material is based upon work supported by the National Science Foundation under grant PHY 1520748.}

\ack{I would like to thank Miranda Cheng, Atish Dabholkar, John Duncan, Greg Moore, Sameer Murthy, Brandon Rayhaun and Yuxiao Wu for many useful conversations on the material discussed here.}

\disclaimer{Any opinions, findings, and conclusions or recommendations expressed in this material are those of the author and do not necessarily reflect the views of the
National Science Foundation.}


\end{document}